\documentclass[12pt]{iopart}

\usepackage{color}
\usepackage[dvipdf]{epsfig}

\usepackage{graphicx}

\begin{document}
\title[Dark Sector from Interacting Canonical and Non-Canonical Scalar Fields ]
{Dark Sector from Interacting Canonical and Non-Canonical Scalar Fields}

\author{Rudinei C. de Souza\dag\ and Gilberto M. Kremer\dag
\footnote[3]{To whom correspondence should be addressed (kremer@fisica.ufpr.br)}
}

\address{\dag\ Departamento de F\'\i sica, Universidade Federal do Paran\'a,
 Curitiba, Brazil}

\def\be{\begin{equation}}
\def\ee#1{\label{#1}\end{equation}}
\newcommand{\ben}{\begin{eqnarray}}
\newcommand{\n}{\nonumber}
\newcommand{\een}{\end{eqnarray}}
\newcommand{\lb}{\label}
\def\v{\varphi}
\def\lb{\label}
\newcommand{\no}{\nonumber}

\begin{abstract}
In this work it is investigated general models with interactions between two
canonical scalar fields and between one non-canonical (tachyon-type)
and one canonical scalar field. The potentials and couplings to the
gravity are selected through the {Noether symmetry approach}.
These general models are employed to describe interactions between {dark energy  and dark matter}, with the fields being
constrained by the astronomical data. The cosmological solutions of
some cases are compared with the observed evolution of the late
Universe.
\end{abstract}

 \date{\today}


\maketitle

\section{Introduction}

It is well known that the {common matter} cannot explain the observed galaxy rotation curves and another
type of matter is naturally necessary. Although this is an old problem \cite{1}, until the present it is not solved and the {most} accepted explanation is that there exists {a strange} matter field which {interacts} only gravitationally with the known matter -- the so-called dark matter \cite{2}. The recent data from the gravitational lensing effect to infer the matter contained in the galaxies strongly supports the existence of the dark matter \cite{3,4}.

More recently, the astronomical observations indicate that the Universe is expanding acceleratedly in the late time \cite{5,6}. But the standard cosmology cannot explain this observed behavior and {cosmologists} are looking for explanations to the current accelerated period. Until the present the most accepted idea is that {there exists} an exotic component with negative pressure which causes the accelerated expansion of the Universe -- the so-called dark energy. It is generally described by a scalar field \cite{7,8} and composes the most part of the energy of the Universe {in the present}.

After the discovery of the accelerated expansion of the Universe, several models taking account {the dark energy and dark matter} -- the so-called dark sector -- were proposed and the most part of them consider {the dark energy and dark matter} as non-interacting fields. More recently, it has appeared models in the literature
where it  was  investigated an interacting dark sector -- an interesting analysis of the viability of {such a} interaction can be found in
\cite{9a} -- and the effects from a possible dark interaction upon the dynamics of galaxy clusters appear to be in
{agreement with} the observations \cite{10a}. In the paper \cite{11a} the authors analyze the energy exchange between
the dark fields. The works \cite{12a,13a,14a} propose models which use \emph{a priori} specified scalar fields
for the representation of the dark sector whereas the works \cite{15a,16a} suppose certain interactions between the
dark fields and represent them by relations {involving} their \emph{a priori} non-specified energy densities
which are posteriorly determined. An interacting dark sector non-minimally coupled to the gravity {is proposed} in
\cite{17a} and in the work \cite{16b} one investigates a model {of dark} energy interacting with
neutrinos and dark matter. The growth of structures under the interaction between dark matter and dark energy was investigated in
the work \cite{18a}. Also in  the matter of scalar fields, the tachyon-type scalar field has received considerable attention in cosmology since {it} can simulate the dark energy with certain success \cite{9,10,10.1,11,12,13,14}.

In order to describe the late Universe, in this work we consider a
spatially flat, homogeneous and isotropic Universe composed by an
interacting dark sector and a {common matter} field. The dark sector
will be investigated from two general models: interacting canonical
scalar fields and interacting non-canonical (tachyon-type) and
canonical scalar fields. The analysis starts from a general action
and the potentials and couplings to the gravity are selected from
the {condition of existence for the Noether symmetry} \cite{14,21,22,23,24}.
{Each set of potentials and couplings satisfying the symmetry
condition corresponds to a particular model. The field equations for
some particular models resulting from the symmetry are solved and
their respective cosmological scenarios are analyzed.} The
cosmological solutions show that these kinds of models {produce}
decelerated-accelerated regimes from a {dynamics} with energy exchange among {the
gravitational field and dark fields}. The resulting cosmological
scenarios present a good agreement with the observational data.

This paper is organized as follows: in the second section the general model of interacting canonical scalar fields is analyzed. {In the} subsection 2.1 the field equations are derived from a point-like Lagrangian. One selects the
potentials and couplings by the Noether symmetry approach in the subsection 2.2. And in the subsection 2.3 the cosmological solutions for the most general cases are obtained.
The third section treats the general model of interacting canonical and non-canonical scalar fields. In the subsection 3.1 the potentials and couplings are selected by the Noether symmetry. The field equations are derived in the section 3.2 (from a point-like Lagrangian) and the resulting equations of energy exchange are obtained. The cosmological solutions for the minimally and non-minimally coupled cases are obtained in the subsection 3.3.
The conclusions about the results close the paper in the fourth section.
In this work we will adopt the signature $(+, -, -, -)$ for the metric and the natural
units $8 \pi G=c=\hbar=1$.

\section{Interacting canonical scalar fields}

\subsection{General action and field equations}

{Let us take} a general action for two interacting scalar fields non-minimally coupled to the gravity of the form
\ben\nonumber
S=\int d^4x\sqrt{-g}\bigg\{[F(\phi)+G(\chi)]R
+\frac{1}{2}\partial_\mu\phi\partial^\mu\phi-V(\phi)\\ +\frac{1}{2}\partial_\mu\chi\partial^\mu\chi-W(\phi,
\chi)\bigg\}\ +\ S_m, \label{ga}
\een
with $S_m=\int d^4x\sqrt{-g}\mathcal{L}_m$ being an additional action which represents a {common matter } field.
Here $R$ is the Ricci scalar {and $F(\phi)$, $G(\chi)$ denote} generic {$C^2$} functions
{which describe the coupling of the scalar fields to the gravitational field}. Furthermore, $V(\phi)$ is the self-interaction potential of the field
$\phi$ and $W(\phi, \chi)$ describes the interaction between the fields $\phi$ and $\chi$, including the
self-interaction of the field $\chi$. In this action the Einstein coupling is recovered when
$F(\phi)+G(\chi)\rightarrow 1/2$.

By varying the action (\ref{ga}) with respect to the metric tensor $g_{\mu\nu}$, we obtain the following modified Einstein's equations
\begin{equation}
R_{\mu\nu}-\frac{1}{2}g_{\mu\nu}R=-\frac{T_{\mu\nu}}{2(F+G)},\label{ee}
\end{equation}
where $T_{\mu\nu}=T_{\mu\nu}^m+T_{\mu\nu}^{\phi}+T_{\mu\nu}^{\chi}$ denotes the total energy-momentum tensor
related to all components of the Universe and the letters $m, \phi$ and $\chi$ designate the energy-momentum
tensor of {the common matter and fields $\phi$ and $\chi$}, respectively. They are given by
\ben
T_{\mu\nu}^m=\frac{2}{\sqrt{-g}}\frac{\delta(\sqrt{-g}\mathcal{L}_m)}{\delta g^{\mu\nu}},
\\
T_{\mu\nu}^\phi=\partial_\mu\phi\partial_\nu\phi-\bigg(\frac{1}{2}\partial_\theta\phi\partial^\theta\phi-
V\bigg)g_{\mu\nu}+2(\nabla_\mu\nabla_\nu-g_{\mu\nu}\nabla_\theta\nabla^\theta)F,
\\
T_{\mu\nu}^\chi=\partial_\mu\chi\partial_\nu\chi-\bigg(\frac{1}{2}\partial_\theta\chi\partial^\theta\chi-
W\bigg)g_{\mu\nu}+2(\nabla_\mu\nabla_\nu-g_{\mu\nu}\nabla_\theta\nabla^\theta)G,
\een
with $\nabla_\mu$ denoting the covariant derivative.

Let us consider a flat FRW metric, $ds^2=dt^2-a(t)^2(dx^2+dy^2+dz^2)$ -- where $a(t)$ is the scale factor -- and suppose that
 the scalar fields are homogeneous, $\phi=\phi(t)$ and $\chi=\chi(t)$, and that the {common matter} is a
pressureless fluid. Hence, we can write from the action (\ref{ga}) the point-like Lagrangian
\ben\nonumber
\mathcal{L}=6a\dot{a}^2(F+G)+6a^2\dot
a\bigg(\frac{dF}{d\phi}\dot\phi+\frac{dG}{d\chi}\dot\chi\bigg)\\-a^3\bigg\{\frac{1}{2}\dot\phi^2-V+
\frac{1}{2}\dot\chi^2-W\bigg\}+\rho_m^0,\label{plL}
\een
where $\rho_m^0$ is the energy density of the {common matter} field {at an initial instant} and the point represents
derivative with respect to time.

The Euler-Lagrange equations applied to the Lagrangian (\ref{plL}) for $a$, $\phi$ and $\chi$ furnish
\ben
2\dot H+3H^2=-\frac{p}{2(F+G)},\label{ae}
\\
\ddot\phi+3H\dot\phi-6(\dot H+2H^2)\frac{dF}{d\phi}+\frac{dV}{d\phi}+\frac{\partial
W}{\partial\phi}=0,\label{phie}
\\
\ddot\chi+3H\dot\chi-6(\dot H+2H^2)\frac{dG}{d\chi}+\frac{\partial W}{\partial\chi}=0,\label{chie}
\een
respectively. Moreover, by imposing that the energy function  associated with the Lagrangian (\ref{plL}) vanishes, one
obtains the modified Friedmann equation, i.e.,
\begin{equation}
E_\mathcal{L}=\frac{\partial \mathcal{L}}{\partial \dot{a}}
\dot{a}+\frac{\partial \mathcal{L}}{\partial\dot{\phi}}\dot{\phi}+\frac{\partial
\mathcal{L}}{\partial\dot{\chi}}\dot{\chi}-\mathcal{L}\equiv0,\; \Longrightarrow\; H^2=\frac{\rho}{6(F+G)}.\label{fe}
\end{equation}
In the above equations $H=\dot a/a$ denotes the Hubble parameter. {The set (\ref{ae})-(\ref{fe}) are the field equations}, where $\rho=\rho_m+\rho_\phi+\rho_\chi$ and $p=p_\phi+p_\chi$, which are given by
\ben
\rho_\phi=\frac{1}{2}\dot\phi^2+V-6H\frac{dF}{d\phi}\dot\phi,\label{phide}
\\
\rho_\chi=\frac{1}{2}\dot\chi^2+W-6H\frac{dG}{d\chi}\dot\chi,\label{chide}
\\
p_\phi=\frac{1}{2}\dot\phi^2-V+2\bigg(\frac{dF}{d\phi}\ddot{\phi}+2H\frac{dF}{d\phi}\dot{\phi}+
\frac{d^2F}{d\phi^2}\dot{\phi}^2\bigg),\label{phip}
\\
p_\chi=\frac{1}{2}\dot\chi^2-W+2\bigg(\frac{dG}{d\chi}\ddot{\chi}+2H\frac{dG}{d\chi}\dot{\chi}+
\frac{d^2G}{d\chi^2}\dot{\chi}^2\bigg).\label{chip}
\een

{Now we will calculate the covariant derivative of the total
energy-momentum tensor, $\nabla_\mu T^{\mu\nu}$ = $\nabla_\mu
T^{\mu\nu}_m+\nabla_\mu T^{\mu\nu}_\phi+\nabla_\mu T^{\mu\nu}_\chi$,
in order to analyze the energy exchange among the fields.
Firstly, the computation of the quantity $\dot\rho+3H(\rho+p)$ for
the energy densities (\ref{phide}) and (\ref{chide}) and their
respective pressures leads to}
\begin{equation}
\dot\rho_{\phi}+3H(\rho_{\phi}+p_{\phi})=-\frac{\partial
W}{\partial\phi}\dot\phi+\frac{(dF/d\phi)\dot\phi}{F+G}\rho,\label{etephi}
\end{equation}
\begin{equation}
\dot\rho_{\chi}+3H(\rho_{\chi}+p_{\chi})=\frac{\partial
W}{\partial\phi}\dot\phi+\frac{(dG/d\chi)\dot\chi}{F+G}\rho,\label{etechi}
\end{equation}
where  (\ref{phie}), (\ref{chie}) and (\ref{fe}) were used
for the simplifications. {The equations} (\ref{etephi}) and
(\ref{etechi}) are the same that those resulting from the covariant derivative
of the energy-momentum tensors for the scalar fields
$\phi$ and $\chi$ ($\nabla_\mu T^{\mu\nu}_\phi$ and $\nabla_\mu
T^{\mu\nu}_\chi$), respectively. Then, remembering that
$\dot\rho_m+3H(\rho_m+p_m)=0$, we have that the covariant derivative
of the total energy-momentum tensor is

\begin{equation}
\nabla_\mu
T^{\mu\nu}=\frac{\rho}{F+G}\bigg(\frac{dF}{d\phi}\dot\phi+\frac{dG}{d\chi}\dot\chi\bigg).\label{temt}
\end{equation}

{By observing (\ref{etephi}) and (\ref{etechi}), we note
that their first terms on the right side represent the energy
exchange between the fields $\phi$ and $\chi$ and their second terms
on the right side describe the energy exchange among {the scalar
fields and gravitational field}. From  (\ref{temt})
one concludes that if $F$ and $G$ are constants, $\nabla_\mu
T^{\mu\nu}=0$, meaning that when the coupling is minimal there is no
energy exchange among {the scalar fields and gravitational field}.
If this is the case, there  exists an energy exchange {only} between the
scalar fields, and consequently the total energy related to the
components of the Universe is conserved.}

\subsection{Couplings and potentials from the Noether symmetry}

{By starting from a general action, we can restrict the forms of the undefined couplings and potentials through the requirement of mathematical proprieties for the Lagrangian, such as symmetries. The symmetries may generate some formal suggestions for the possible forms of the undefined functions. In this work we will require that the Lagrangian of the general model satisfies the Noether symmetry, which provides a conserved quantity associated with the dynamical system. Interesting results may arise from the Noether symmetry approach, as can be seen in the works \cite{14,21,22,23,24}}.

{A Noether symmetry for a given Lagrangian of the form $\mathcal{L}=\mathcal{L}(q_i, \dot q_i)$ exists if the condition} $L_{x}\mathcal{L}={X}\mathcal{L}=0$ {is satisfied, with $L_{x}$ designating the Lie derivative with respect to the vector field ${X}$
defined by}
\begin{equation}
{X}=\alpha_i\frac{\partial}{\partial q_i}+\frac{d\alpha_i}{dt}\frac{\partial}{\partial\dot q_i},
\end{equation}
{where the $\alpha_i$'s are functions of the generalized coordinates $q_i$. The constant of motion associated with the Noether symmetry generated by ${X}$ is given by
}
\begin{equation}
M_0=\alpha_i\frac{\partial\mathcal{L}}{\partial \dot q_i}. {\label{constmot}}
\end{equation}

The condition of existence for the {Noether symmetry $L_{{x}}\mathcal{L}={X}\mathcal{L}=0$ is } applied to the
point-like Lagrangian ({\ref{plL}}), with the vector field ${X}$ defined for our problem as follows
\be
{X}=\alpha\frac{\partial}{\partial a}+\beta\frac{\partial}{\partial
\phi}+\gamma\frac{\partial}{\partial \chi}+\frac{\partial\alpha}{\partial t}\frac{\partial}{\partial
\dot{a}}+\frac{\partial \beta}{\partial t}\frac{\partial}{\partial
\dot{\phi}}+\frac{\partial \gamma}{\partial t}\frac{\partial}{\partial \dot{\chi}},
\ee{X}
where $\alpha$, $\beta$ and $\gamma$ are functions of ($a$, $\phi$, $\chi$). In this case we obtain the following coupled system
of differential equations
\ben
(F+G)\bigg(\alpha+2a\frac{\partial \alpha}{\partial a}\bigg)+a\frac{dF}{d\phi}\bigg(\beta+a\frac{\partial
\beta}{\partial a} \bigg)+a\frac{dG}{d\chi}\bigg(\gamma+a\frac{\partial \gamma}{\partial a} \bigg)=0,\label{1s}
\\
3\alpha-12\frac{dF}{d\phi}\frac{\partial\alpha}{\partial\phi}+2a\frac{\partial\beta}{\partial\phi}=0,\label{2s}
\\
3\alpha-12\frac{dG}{d\chi}\frac{\partial\alpha}{\partial\chi}+2a\frac{\partial\gamma}{\partial\chi}=0,\label{3s}
\\\nonumber
a\beta\frac{d^2F}{d\phi^2}+\bigg(2\alpha+a\frac{\partial\alpha}{\partial
a}+a\frac{\partial\beta}{\partial\phi}\bigg)\frac{dF}{d\phi}+a\frac{\partial\gamma}{\partial\phi}\frac{dG}{d\chi}
\\+2\frac{\partial\alpha}{\partial\phi}(F+G)-\frac{a^2}{6}\frac{\partial\beta}{\partial a}=0,\label{4s}
\\\nonumber
a\gamma\frac{d^2G}{d\chi^2}+\bigg(2\alpha+a\frac{\partial\alpha}{\partial
a}+a\frac{\partial\gamma}{\partial\chi}\bigg)\frac{dG}{d\chi}+a\frac{\partial\beta}{\partial\chi}\frac{dF}{d\phi}
\\+2\frac{\partial\alpha}{\partial\chi}(F+G)-\frac{a^2}{6}\frac{\partial\gamma}{\partial a}=0,\label{5s}
\een
\ben
\frac{\partial\alpha}{\partial\phi}\frac{dG}{d\chi}+\frac{\partial\alpha}{\partial\chi}\frac{dF}{d\phi}
-\frac{a}{6}\bigg(\frac{\partial\beta}{\partial\chi}+\frac{\partial\gamma}{\partial\phi}\bigg)=0,\label{6s}
\\
3\alpha(V+W)+a\beta\bigg(\frac{dV}{d\phi}+\frac{\partial W}{\partial\phi}\bigg)+a\gamma\frac{\partial
W}{\partial\chi}=0.\label{7s}
\een

The solution of the coupled system of differential equations (\ref{1s})-(\ref{7s}) is not unique and {the several solutions} are found {in Tables 1 and 2} which contain all sets of functions $\alpha, \beta, \gamma, F, G, V, W$, where the quantities $\alpha_0, \beta_0, \gamma_0, F_0, F_0^1,$ $G_0, G_0^1, V_0, W_0$ are
constants and $K=\beta_0/\gamma_0$. We have looked for {solutions} which always furnish {for the function $W$ an expression different} from (0, constant,
$f(\phi)$,  $g(\chi)$) in order to guarantee an interaction between the fields $\phi$ and $\chi$.

\begin{table}[h]
\centering
\begin{tabular}{|c|c|c|c|c|c|c|c|}
\hline
 & $\alpha$ & $\beta$ & $\gamma$ & $F$ & $G$ & $V$ & $W$ \\
\hline
I & $\alpha_0a$ & $-3\alpha_0\phi/2$ & $-3\alpha_0\chi/2$ & $F_0\phi^2$ & $G_0\chi^2$ & 0, $V_0\phi^2$ & $f(\chi/\phi)\phi^2$  \\
\hline
II & $\alpha_0a$ & $-3\alpha_0\phi/2$ & $-3\alpha_0\chi/2$ & $F_0\phi^2$ & $0$ & 0, $V_0\phi^2$ & $f(\chi/\phi)\phi^2$  \\
\hline
III & $\alpha_0a$ & $-3\alpha_0\phi/2$ & $-3\alpha_0\chi/2$ & $0$ & $G_0\chi^2$ & 0, $V_0\phi^2$ & $f(\chi/\phi)\phi^2$  \\
\hline
\end{tabular}
\caption{Solutions with $(\alpha, \beta, \gamma)\neq0$.}
\end{table}

One may observe from {Table 2} that the general forms of $W$ provided by the Noether symmetry allow the existence of
sums which incorporate terms of the form $f(\phi)$ -- representing an additional term of self-interaction for
the field $\phi$ -- namely, $W=f(\phi)+g(\phi, \chi)$. In this  case, one must redefine the potentials
in the energy density equations (\ref{phide}) and (\ref{chide}) {and pressure equations} (\ref{phip}) and (\ref{chip}) by writing $W\rightarrow\overline W=W-f(\phi)$ and $V\rightarrow\overline V=V+f(\phi)$. {So we take account the additional self-interaction term of the field $\phi$}.

\begin{table}[h]
\centering
\begin{tabular}{|c|c|c|c|c|c|c|}
\hline
 & $\beta$ & $\gamma$ & $F$ & $G$ & $V$ & $W$ \\
\hline I & $\beta_0\chi$ & $-\beta_0\phi$ & $F_0^1+F_0\phi^2$ &
$G_0^1+G_0\chi^2$ & $\int(\frac{\phi}{\chi}\frac{\partial
W}{\partial\chi}
-\frac{\partial W}{\partial\phi})d\phi$ & $W_0\int h(\phi^2+\chi^2)\chi d\chi$  \\
\hline II & $\beta_0\chi$ & $-\beta_0\phi$ & $F_0^1+F_0\phi^2$ &
$G_0^1+G_0\chi^2$ &
0, $V_0$ & $W(\phi^2+\chi^2)$  \\
\hline III & $\beta_0\chi$ & $-\beta_0\phi$ & $F_0$ & $G_0$ &
$\int(\frac{\phi}{\chi}\frac{\partial W}{\partial\chi}
-\frac{\partial W}{\partial\phi})d\phi$ & $W_0\int h(\phi^2+\chi^2)\chi d\chi$  \\
\hline IV & $\beta_0\chi$ & $-\beta_0\phi$ & $F_0$ & $G_0$ &
0, $V_0$ & $W(\phi^2+\chi^2)$  \\
\hline V & $\beta_0$ & {$\gamma_0$} & $F_0^1+F_0\phi$ &
$G_0^1-KF_0\chi$ &
0, $V_0$ & $W(\phi-K\chi)$  \\
\hline VI & $\beta_0$ & $\gamma_0$ & $F_0$ & $G_0$ &
0, $V_0$ & $W(\phi-K\chi)$  \\
\hline
\end{tabular}
\caption{Solutions with $\alpha=0$ and $(\beta, \gamma)\neq0$.}
\end{table}

\begin{table}
\centering
\begin{tabular}{|c|c|}
\hline
Cases & $M_0$ \\
\hline
 I--III Tab 1& $\frac{3}{2}\alpha_0a^3\bigg\{2H\Big[4(F+G)-3\Big(\phi\frac{dF}{d\phi}+\chi\frac{dG}{d\chi}\Big)\Big]
+\Big(4\frac{dF}{d\phi}+\phi\Big)\dot\phi+\Big(4\frac{dG}{d\chi}+\chi\Big)\dot\chi\bigg\}$  \\
\hline
 I--IV Tab 2 &
  $\beta_0a^3\bigg\{6H\Big(\chi\frac{dF}{d\phi}-\phi\frac{dG}{d\chi}\Big)+\phi\dot\chi-\chi\dot\phi\bigg\}$  \\
\hline
V--VI Tab 2 &  $\gamma_0a^3\bigg\{6H\Big(K\frac{dF}{d\phi}+\frac{dG}{d\chi}\Big)-K\dot\phi-\dot\chi\bigg\}$  \\
\hline
\end{tabular}
\caption{Constants of motion.}
\end{table}

The case II {in Table} 1 with $V=V_0\phi^2$ is similar to the model analyzed in the work \cite{17a} and the
{case} III {in Table} 2, when $h=\phi^2+\chi^2$, is the model proposed in \cite{12a} with $V=W_0\phi^4$ but with
an additional self-interaction term of the form $W_0\chi^4$. The models of the works \cite{12a,17a} are
particular cases of the one denoted by I {in Table} 2.

{From the equation (\ref{constmot}) we can write the constants of motion associated with the cases in Tables 1 and 2. They are summarized in Table 3.}

\subsection{Solutions of the field equations}

Due to the interaction between the fields $\phi$ and $\chi$ and the presence of a {common matter} field in the action
(\ref{ga}), the field equations become more complicated than in the case with two non-interacting scalar fields
and without a {common matter field \cite{22}}.
Hence, the search for numerical solutions of the system (\ref{ae})-(\ref{fe}) for the most general cases {in Tables 1
and 2 will be performed}.

Let us transform the derivatives with respect to time in {the equations }(\ref{ae})-(\ref{fe}) into derivatives with
respect {to red-shift} through the relationships
\begin{equation}
z=\frac{1}{a}-1,\qquad \frac{d}{dt}=-H(1+z)\frac{d}{dz},
\end{equation}
and divide all {the equations} by $\rho_0$ -- the total energy density of the Universe at the present time. Hence, one obtains from the equations (\ref{ae})-(\ref{fe})
the following system of coupled differential equations
\ben
4 \widetilde{H}\widetilde{H}'(1+z)(F+G)=\widetilde{\rho}+\widetilde{p},\label{ae1}
\\\nonumber
\widetilde{H}^2(1 +
z)^2\phi''+\widetilde{H}\left[\widetilde{H}'(1+z)-2\widetilde{H}\right]\left[(1+z)\phi'+6\frac{dF}{d\phi}\right]\\+
\frac{d\widetilde{V}}{d\phi}+\frac{\partial\widetilde{W}}{\partial\phi}=0,\label{phie1}
\\
\widetilde{H}^2(1 + z)^2\chi''+\widetilde{H}\left[\widetilde{H}'(1 +
z)-2\widetilde{H}\right]\left[(1+z)\chi'+6\frac{dG}{d\chi}\right]+\frac{\partial\widetilde{W}}{\partial\chi}=0,\label{chie1}
\een
by taking into account the equation  (\ref{fe}). Above, the prime represents the  derivative with respect to $z$
and the following dimensionless quantities were introduced:
$\widetilde{\rho}=\rho/\rho_0=\rho_m/\rho_0+\rho_\phi/\rho_0+\rho_\chi/\rho_0=\widetilde{\rho}_m+
\widetilde{\rho}_\phi+\widetilde{\rho}_\chi,$
$ \widetilde{p}=p/\rho_0=p_\phi/\rho_0+p_\chi/\rho_0=\widetilde{p}_\phi+\widetilde{p}_\chi,$
$\widetilde{H}=H/\sqrt{\rho_0},$ $\widetilde{V}=V/\rho_0,$ and  $\widetilde{W}=W/\rho_0.$
Furthermore,  the dimensionless energy densities and pressures read
\ben
\widetilde{\rho}_m=\widetilde{\rho}_m^0(1 + z)^3,\label{edm1}
\\
\widetilde{\rho}_{\phi}=\frac{\widetilde{H}^2(1 +
z)^2\phi'^2}{2}+\widetilde{V}+6\widetilde{H}^2(1 + z)\frac{dF}{d\phi}\phi',\label{edphi1}
\\
\widetilde{\rho}_{\chi}=\frac{\widetilde{H}^2(1 +
z)^2\chi'^2}{2}+\widetilde{W}+6\widetilde{H}^2(1 + z)\frac{dG}{d\chi}\chi',\label{edchi1}
\\\nonumber
\widetilde{p}_{\phi}=\frac{\widetilde{H}^2(1 +
z)^2\phi'^2}{2}-\widetilde{V}+2\widetilde{H}(1+z)\bigg\{\widetilde{H}(1+z)\bigg(\frac{d^2F}{d\phi^2}\phi'^2+
\frac{dF}{d\phi}\phi''\bigg)\\
+ \ [\widetilde{H}'(1+z)-\widetilde{H}]\frac{dF}{d\phi}\phi'\bigg\},\label{pphi1}
\\\nonumber
\widetilde{p}_{\chi}=\frac{\widetilde{H}^2(1 +
z)^2\chi'^2}{2}-\widetilde{W}+2\widetilde{H}(1+z)\bigg\{\widetilde{H}(1+z)\bigg(\frac{d^2G}{d\chi^2}\chi'^2+
\frac{dG}{d\chi}\chi''\bigg)\\
+ \ [\widetilde{H}'(1+z)-\widetilde{H}]\frac{dG}{d\chi}\chi'\bigg\}.\label{pchi1}
\een

Our aim is to use the solutions given {in Tables} 1 and 2 in order to describe the dark sector as an interacting structure.
 Then we consider that the fields $\phi$ and $\chi$ correspond to the dark energy and dark matter fields, respectively. For this choice we have to require
certain features for each field: (i) the field $\phi$ must have a negative pressure in the late time and its
energy density composes the most part of the total energy density of the Universe at the present time;
(ii) the field $\chi$ has a small positive pressure in comparison to the pressure modulus of the dark energy and
its energy density still represents a considerable fraction of the total energy density of the Universe at the present time.

To satisfy the above requirements we will use the initial conditions for the system (\ref{ae1})-(\ref{chie1}) which match  the
 astronomical data. At $z=0$ one introduces the quantities $\widetilde{\rho}_m(0)=\rho_m^0/\rho_0=\Omega_m^0$,
$\widetilde{\rho}_\phi(0)=\rho_\phi^0/\rho_0=\Omega_\phi^0$ and
$\widetilde{\rho}_\chi(0)=\rho_\chi^0/\rho_0=\Omega_\chi^0$, where $\Omega^0_i$ denotes the value of the
  density parameter of each component at the present time whereas $\Omega_0=\Omega_m^0+\Omega_\phi^0+\Omega_\chi^0$
refers to the total density parameter. The values of the density parameters adopted here are:
$\Omega_m^0=0.05$, $\Omega_\phi^0=0.72$ and $\Omega_\chi^0=0.23$ (see e.g. reference \cite{25}). Further, in
agreement with the requirement (i) one has that $\phi'(0)^2\ll1$, which means that the field $\phi$ varies very
slowly in the late time, {i.e., $\phi'(0)=\epsilon$, with $\epsilon$ very small.} From (\ref{edphi1}) and this last condition it follows that
$\widetilde{V}(0)\approx\Omega_\phi^0$ and one may obtain the initial condition for $\phi$. Once
$\phi(0)$ is fixed, one may determine the initial condition for $\chi$ from the necessity that the coupling has the present
value 1/2, i.e., $F(0)+G(0)=1/2$.
{Satisfying the requirement II by the condition $\widetilde{\rho}_\chi(0)=\Omega_\chi^0$, and since one
knows $\chi(0)$, from (\ref{edchi1}) we may determine the initial condition for $\chi'$}. The part of the
requirement (ii) that is  related to the value of $p_\chi$ can be satisfied through the adjustments of the constants
that appear in the functions of the couplings and potentials. From (\ref{fe}) we have for the
Hubble parameter the initial condition $\widetilde{H}(0)=\sqrt{\Omega_0/6[F(0)+G(0)]}=1/\sqrt{3}$.
To sum up we have:
\begin{itemize}
\item[(a)]$\widetilde{H}(0) = \frac{1}{\sqrt{3}};$
\item[(b)]$\widetilde{V}(0)=\Omega_\phi^0$  {determines} $\phi(0),\qquad \phi'(0)^2 \ \ll \ 1$;
\item[(c)] $G(0)=\frac{1}{2}-F(0)$ {determines} $\chi(0)$;
\item[(d)]$\chi'(0)^2+6\widetilde{W}(0)+12\chi'(0)\frac{dG}{d\chi}\bigg|_{z=0}=6\Omega_\chi^0$
{determines} $\chi'(0)$.
\end{itemize}

{For the case I {in Table} 1 we take
\begin{equation}\nonumber V=V_0\phi^2, \qquad
f\Big(\frac{\chi}{\phi}\Big)=\frac{\chi}{\phi} \quad \hbox{which
implies} \quad W=W_0\phi\chi.
\end{equation}
By using the above equations, the initial conditions are:
\ben\nonumber
\phi(0)=\sqrt{\frac{0.72}{\widetilde{V_0}}},\qquad
\chi(0)=\sqrt{\frac{1/2-F_0\phi(0)^2}{G_0}}, \\
\chi'(0)=\sqrt{6\big[0.23
+24G_0^2\chi(0)^2-\widetilde{W_0}\phi(0)\chi(0)\big]}-12G_0\chi(0),
\nonumber
\een
where
\begin{equation}\nonumber
F_0\leq\frac{1}{2\phi(0)^2} \quad \hbox{and} \quad
\widetilde{W}_0\leq\frac{0.23/\chi(0)+24G_0^2\chi(0)}{\phi(0)}.
\end{equation}
}
For this case we have adopted the following values for the fixed
constants in the numerical computations: $F_0 = -0.002366, G_0 =
0.04651, V_0 = 0.01001$ and $W_0 = 0.01856$.

{Now, for the case I {in Table} 2, by considering $F_0^1=G_0^1=0$ without
loss of generality, one takes
\ben
\nonumber \ h(\phi^2+\chi^2)=\phi^2+\chi^2 \;\, \hbox{which implies}\;\,
 W=W_0(\chi^4+2\phi^2\chi^2),\; V=W_0\phi^4,
\een
with $W_0$ from {Table 2} replaced by $4W_0$.}

{For this case one has the initial conditions:
\ben
\nonumber \phi(0)=\bigg(\frac{0.72}{\widetilde{W_0}}\bigg)^{1/4},
\qquad
\chi(0)=\sqrt{\frac{1/2-F_0\phi(0)^2}{G_0}}, \nonumber \\
 \chi'(0)=\sqrt{6\big[0.23+(24G_0^2-\widetilde{W_0})\chi(0)^4-2\widetilde{W_0}\phi(0)^2\chi(0)^2\big]}\nonumber  -12G_0\chi(0),
\nonumber
\een
where
\begin{equation}\nonumber F_0\leq\frac{1}{2\phi(0)^2} \quad \hbox{and}\quad
\widetilde{W}_0\leq\frac{0.23+24G_0^2\chi(0)^4}{\chi(0)^4+2\phi(0)^2\chi(0)^2}.
\end{equation}}
 In this case we have taken for the fixed constants the values:
$F_0 = 0.2064,$ $G_0 = 0.03333,$ and  $W_0 = 0.1250.$

\begin{figure}
 \begin{center}
 \vskip0.5cm
 \includegraphics[height=4.6cm,width=6.6cm]{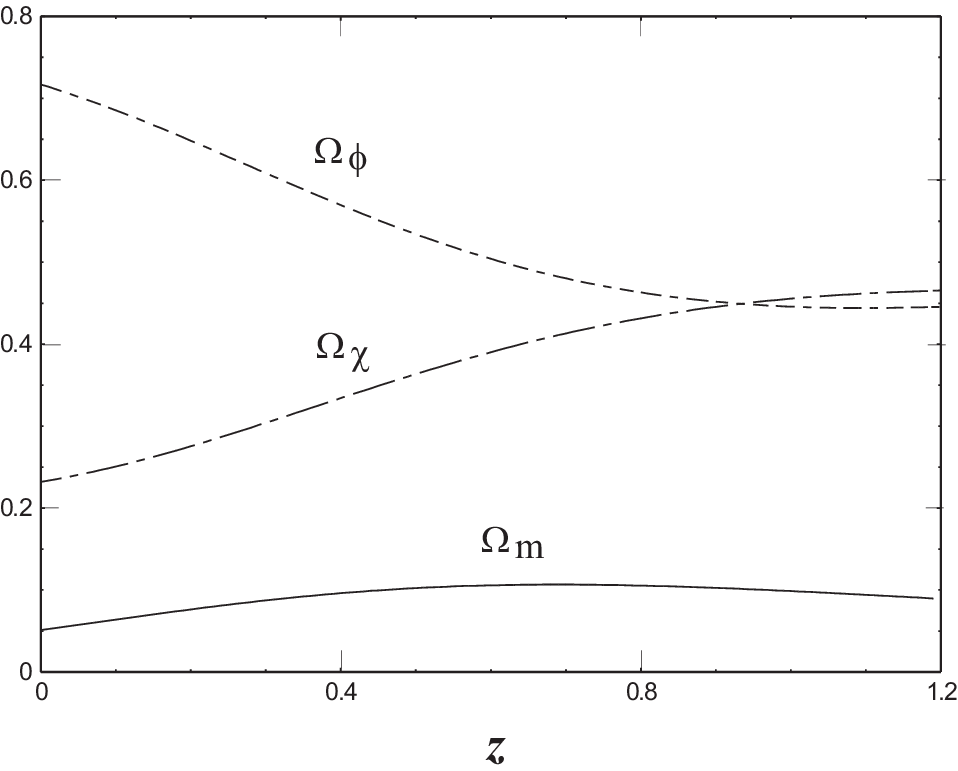}\hskip0.5cm
 \includegraphics[height=4.6cm,width=6.6cm]{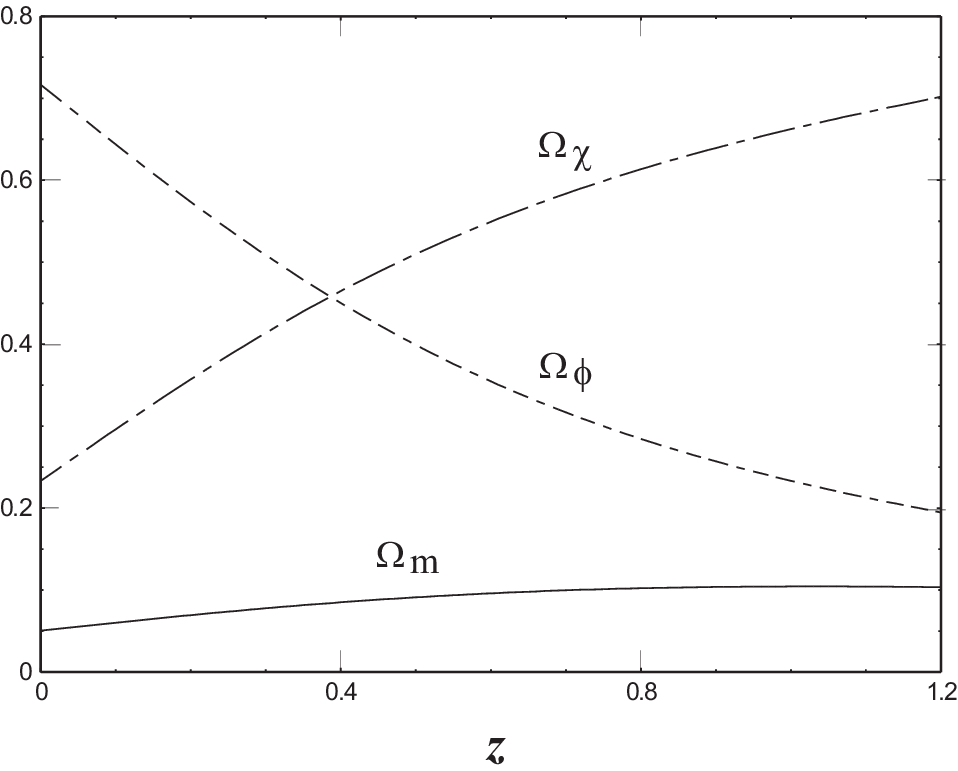}
  \caption{Density parameters {of the common matter and scalar fields} as functions of the
  red-shift $z$. Left frame: case 1; right frame: case 2. \label{f1}.}
 \end{center}
 \end{figure}

In  Figure 1 are represented the density parameters of {the common matter}, dark energy and dark matter for the cases
1 and 2 in the left   and right frames, respectively. From this figure we can observe the evident
difference   of the increase of the density parameter of the quintessence with the red-shift and the corresponding decrease of the
density parameter of the dark matter for the two cases. Since the gravitational coupling has quadratic forms in
both cases, the different red-shift evolutions of the density parameters are determined uniquely by the
interaction and self-interaction potentials of the fields, which are the responsible of the energy transfer  between the scalar fields and among {the scalar fields and gravitational field}, as can be observed from
(\ref{etephi}) and (\ref{etechi}). Then these energy transfer   among the fields (scalar field - scalar field and gravitational field - scalar
fields) have a definitive role in the variety of behaviors which can be produced by models with scalar fields,
as can be seen from these two cases in  Figure 1. This is definitely verified when one observes from
Figure 1 that the {common matter} field, which is not coupled to the other fields, has a red-shift evolution of its
density parameter quite identical in the two cases, meaning that it is just submitted to the dilution caused by
the expansion of the Universe, which presents practically the same rate for the two cases. From these
proprieties, an interacting dark sector could present more possibilities for the energy density evolution of the dark matter.

\begin{figure}
 \begin{center}
 \vskip0.5cm
 \includegraphics[height=4.6cm,width=6.6cm]{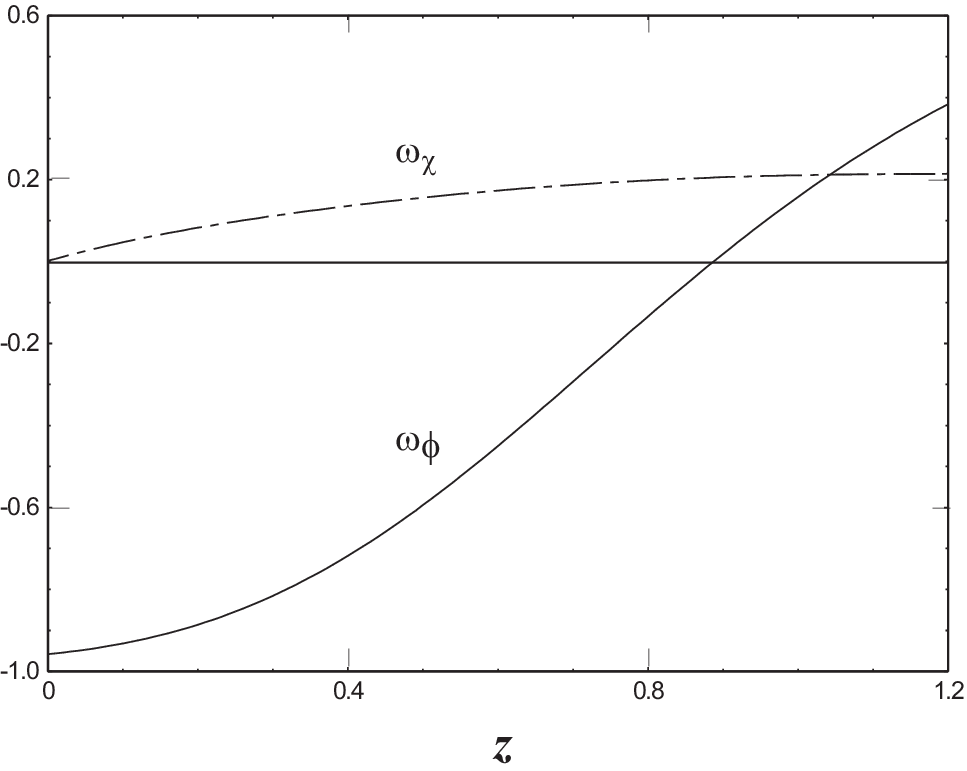}\hskip0.5cm
 \includegraphics[height=4.6cm,width=6.6cm]{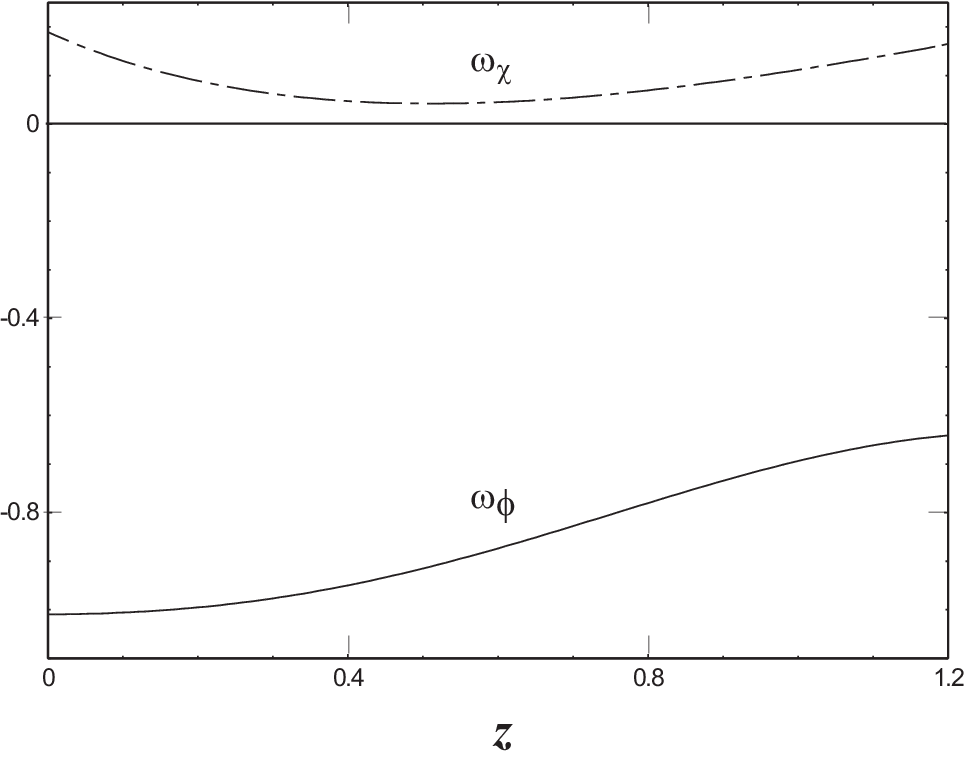}
  \caption{Ratio of the pressure and energy density of the scalar fields as functions
  of the red-shift $z$. Left frame: case 1; right frame:  case 2. \label{f2}}
 \end{center}
 \end{figure}

The ratio of the pressure and energy density of the scalar fields $\omega_\phi=p_\phi/\rho_\phi$ and
$\omega_\chi=p_\chi/\rho_\chi$ are plotted in Figure 2, where the left frame corresponds to the case 1 and the right frame
the case 2. These figures show that the pressure relative to the energy density of the dark matter field is small
in comparison to the one (in modulus) of the dark energy for both cases. But the dark
matter pressure has a significant role in the determination of the epoch where the transition of a
decelerated to an accelerated expansion of the Universe occurs, since its participation in the total composition of the
Universe is significant. Observe that in the present day $\omega_\phi\rightarrow-1$, which corresponds to a
cosmological constant.

\begin{figure}
 \begin{center}
 \vskip0.5cm
 \includegraphics[height=4.7cm,width=6.7cm]{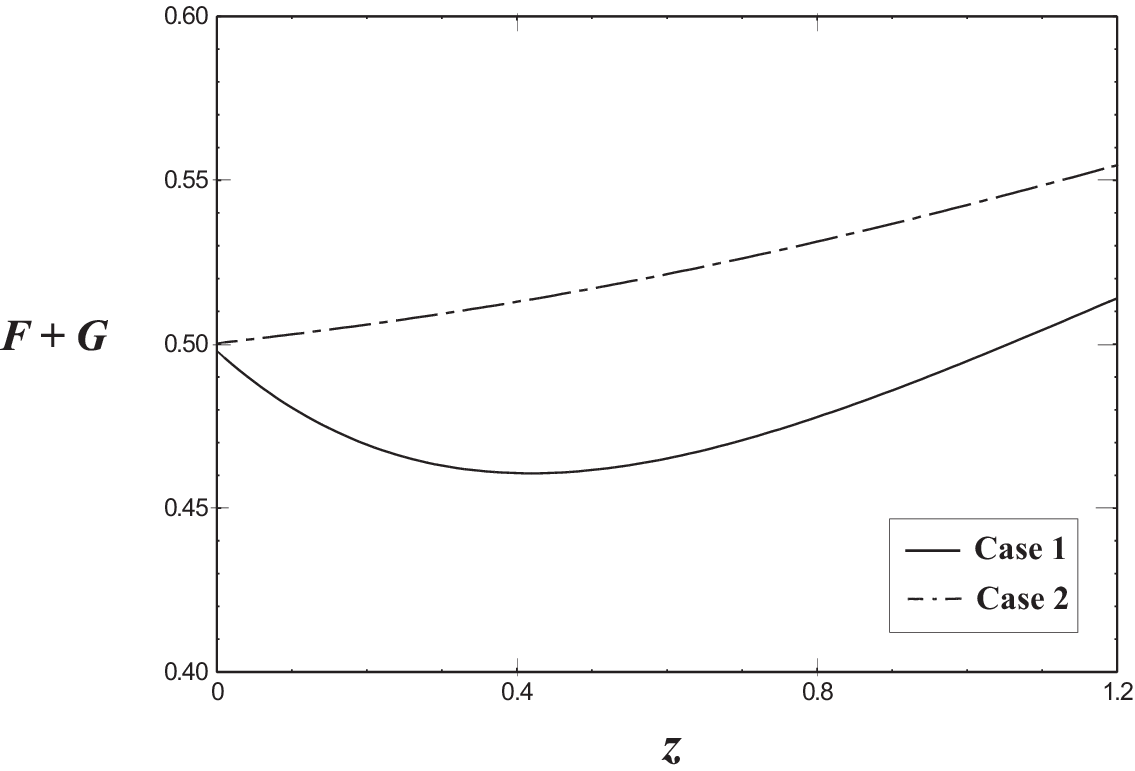}\hskip0.5cm
 \includegraphics[height=4.7cm,width=6.5cm]{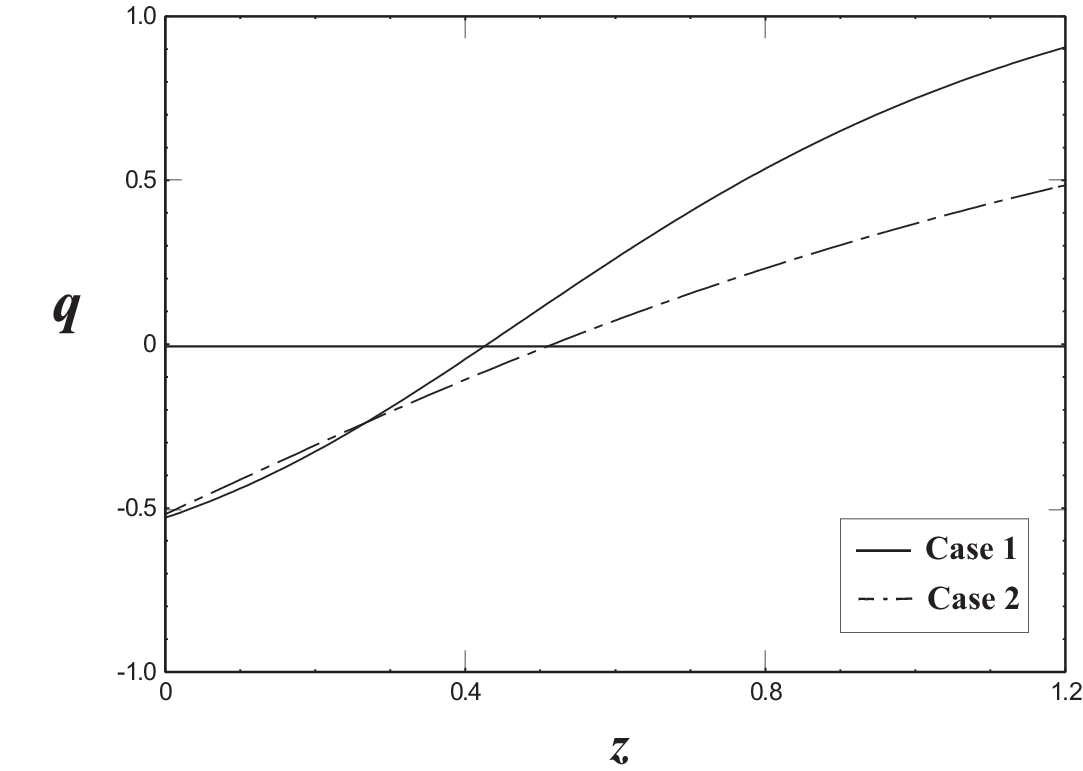}
  \caption{Left frame: effective coupling versus red-shift $z$;
  right frame: deceleration parameter versus red-shift $z$. The straight line represents  the case 1 and the dashed line  the case 2. \label{f3}}
 \end{center}
 \end{figure}

In the left frame of  Figure 3 it is plotted the effective coupling $F+G$ and in the right frame the
deceleration parameter $q = \frac{1}{2}+\frac{3p}{2\rho}$, for the cases 1 and 2. {We can infer from  Figure 3 that the variation of the effective gravitational coupling presents a small value around its present value $F(0)+G(0)=1/2$. This variation is about 10 percent, meaning that the
effective gravitational {"constant"} varies approximately 10 percent in the considered interval.} The right frame of Figure 3 shows that there exists a small difference between the red-shifts of the transition from a decelerated to an
accelerated regime for the two cases. For the cases 1 and 2 the values of the red-shift transitions
are $z_T = 0.43$ and $z_T = 0.52$ whereas the present values of the
deceleration parameter read $q(0) = -0.53$ and $q(0) = -0.52$, respectively. {These results are in good agreement with the observational data, namely,
  $z_T = 0.74 \pm 0.18$ (from \cite{26}) and $q(0) = -0.46 \pm 0.13$ (from \cite{27}).}

\section{Interacting canonical and non-canonical scalar fields}

\subsection{General action and Noether symmetry}

Now let us take an action where one scalar field is non-canonical
and represented by $\varphi$, being a tachyon-type field, and the
other is a canonical scalar field represented by $\phi$
\ben\no
S&=&\int d^4x\sqrt{-g}\bigg\{[F(\varphi)+G(\phi)]R
-V(\varphi)\sqrt{1-\partial_\mu\varphi\partial^\mu\varphi}\\
&+&\frac{1}{2}\partial_\mu\phi\partial^\mu\phi-W(\varphi,
\phi)\bigg\}\ +\ S_m, \label{gat}
\een
where $F(\varphi)$ and $G(\phi)$ represent generic {$C^2$}
functions {which describe the coupling of the scalar fields to the gravity}, $V(\varphi)$ is the self-interaction potential of the
field $\varphi$ and $W(\varphi, \phi)$ describes the
self-interaction of the field $\phi$ and the interaction {between the fields} $\varphi$ and $\phi$. As before, when
$F(\varphi)+G(\phi)\rightarrow 1/2$ we recover the Einstein
coupling.

From the variation of the action (\ref{gat}) with respect to
$g_{\mu\nu}$ one obtains the modified Einstein's equations with the
same form of (\ref{ee})
\begin{equation}
R_{\mu\nu}-\frac{1}{2}g_{\mu\nu}R=-\frac{T_{\mu\nu}}{2(F+G)},\label{eet}
\end{equation}
being
$T_{\mu\nu}=T_{\mu\nu}^m+T_{\mu\nu}^{\varphi}+T_{\mu\nu}^{\phi}$
defined as follows
\ben
T_{\mu\nu}^m=\frac{2}{\sqrt{-g}}\frac{\delta(\sqrt{-g}\mathcal{L}_m)}{\delta
g^{\mu\nu}},\label{et1}
\\
T_{\mu\nu}^\varphi=V\bigg(\frac{\partial_\mu\varphi\partial_\nu\varphi}{\sqrt{1-\partial_\theta\varphi\partial^\theta\varphi}}+
g_{\mu\nu}\sqrt{1-\partial_\theta\varphi\partial^\theta\varphi}\bigg)+2(\nabla_\mu\nabla_\nu-g_{\mu\nu}\nabla_\theta\nabla^\theta)F,\nonumber\\
\label{et2}
\\
T_{\mu\nu}^\phi=\partial_\mu\phi\partial_\nu\phi-\bigg(\frac{1}{2}\partial_\theta\phi\partial^\theta\phi-
W\bigg)g_{\mu\nu}+2(\nabla_\mu\nabla_\nu-g_{\mu\nu}\nabla_\theta\nabla^\theta)G.\label{et3}
\een

By considering again a flat FRW metric and the scalar fields
homogeneous, $\varphi=\varphi(t)$ and $\phi=\phi(t)$, with the
{common matter} being a pressureless fluid, the point-like Lagrangian which follows from the action
(\ref{gat}) reads
\ben\no
\mathcal{L}&=&6a\dot{a}^2(F+G)+6a^2\dot
a\bigg(\frac{dF}{d\varphi}\dot\varphi+\frac{dG}{d\phi}\dot\phi\bigg)\\&+&a^3V\sqrt{1-\dot\varphi^2}
-a^3\bigg(\frac{1}{2}\dot\phi^2-W\bigg)+\rho_m^0.\label{plLt}
\een

{From the condition of existence for the
Noether symmetry $L_{{x}}\mathcal{L}={X}\mathcal{L}=0$
applied to the point-like Lagrangian }(\ref{plLt}), with the vector
field ${X}$ now defined as
\be
{X}=\alpha\frac{\partial}{\partial a}+\beta\frac{\partial}{\partial
\varphi}+\gamma\frac{\partial}{\partial
\phi}+\frac{\partial\alpha}{\partial t}\frac{\partial}{\partial
\dot{a}}+\frac{\partial \beta}{\partial t}\frac{\partial}{\partial
\dot{\varphi}}+\frac{\partial \gamma}{\partial
t}\frac{\partial}{\partial \dot{\phi}},
\ee{a}
where $\alpha$, $\beta$ and $\gamma$ are functions of ($a$, $\varphi$,
$\phi$), we obtain the following system of partial differential
equations
\ben
(F+G)\bigg(\alpha+2a\frac{\partial \alpha}{\partial
a}\bigg)+a\frac{dF}{d\varphi}\bigg(\beta+a\frac{\partial
\beta}{\partial a}
\bigg)
+a\frac{dG}{d\phi}\bigg(\gamma+a\frac{\partial
\gamma}{\partial a} \bigg)=0,\label{1st}
\\
\frac{\partial \alpha}{\partial \varphi}{dF\over d\varphi}=0, \qquad
\frac{\partial \beta}{\partial a}=0, \qquad
\frac{\partial\beta}{\partial \varphi}=0, \qquad
\frac{\partial\beta}{\partial \phi}=0,\label{2st}
\\
3\alpha-12\frac{dG}{d\phi}\frac{\partial\alpha}{\partial\phi}+2a\frac{\partial\gamma}{\partial\phi}=0,\label{3st}
\\\no
a\beta\frac{d^2F}{d\varphi^2}+\bigg(2\alpha+a\frac{\partial\alpha}{\partial
a}+a\frac{\partial\beta}{\partial\varphi}\bigg)\frac{dF}{d\varphi}\\+a\frac{\partial\gamma}{\partial\varphi}\frac{dG}{d\phi}
+2\frac{\partial\alpha}{\partial\varphi}(F+G)=0,\label{4st}
\\\no
a\gamma\frac{d^2G}{d\phi^2}+\bigg(2\alpha+a\frac{\partial\alpha}{\partial
a}+a\frac{\partial\gamma}{\partial\phi}\bigg)\frac{dG}{d\phi}+a\frac{\partial\beta}{\partial\phi}\frac{dF}{d\varphi}
\\+2\frac{\partial\alpha}{\partial\phi}(F+G)-\frac{a^2}{6}\frac{\partial\gamma}{\partial
a}=0,\label{5st}
\\
\frac{\partial\alpha}{\partial\varphi}\frac{dG}{d\phi}+\frac{\partial\alpha}{\partial\phi}\frac{dF}{d\varphi}
-\frac{a}{6}\bigg(\frac{\partial\beta}{\partial\phi}+\frac{\partial\gamma}{\partial\varphi}\bigg)=0,\label{6st}
\\
3\alpha V+a\beta\frac{dV}{d\varphi}=0,\label{7st}
\\
3\alpha W+a\beta\frac{\partial
W}{\partial\varphi}+a\gamma\frac{\partial
W}{\partial\phi}=0.\label{8st}
\een

The solution of the system (\ref{1st})-(\ref{8st}) is not unique and
the solutions that we found are given {in Table} 4 containing the sets
of $\alpha, \beta, \gamma, F, G, V, W$, where $\alpha_0, \beta_0,
\gamma_0, F_0, F_0^1,$ $G_0, G_0^1$ and $V_0$ are constants and
{$\mu=3\alpha_0/\beta_0$} and $K=\beta_0/\gamma_0$. Here we also
looked for solutions which present potentials of the form  $W\neq
(0$, constant, $f(\varphi), g(\phi))$ in order to provide an
interaction between the fields $\varphi$ and $\phi$. {And from (\ref{constmot})
we have the respective constants of motion, which are given in Table 5.}

It is interesting to observe that the solution I generalizes the
model analyzed in the work \cite{14}.

\begin{table}
\centering
\begin{tabular}{|c|c|c|c|c|c|c|c|}
\hline
 & $\alpha$ & $\beta$ & $\gamma$ & $F$ & $G$ & $V$ & $W$ \\
\hline
I & $\alpha_0a$ & $\beta_0$ & $-3\alpha_0\phi/2$ & $F_0e^{-\mu\varphi}$ & $G_0\phi^2$ & $V_0e^{-\mu\varphi}$ & $f\big(\phi e^{\frac{\mu\varphi}{2}}\big)e^{-\mu\varphi}$  \\
\hline
II & $\alpha_0a$ & $\beta_0$ & $-3\alpha_0\phi/2$ & $F_0e^{-\mu\varphi}$ & $0$ & $V_0e^{-\mu\varphi}$ & $f\big(\phi e^{\frac{\mu\varphi}{2}}\big)e^{-\mu\varphi}$  \\
\hline
III & $\alpha_0a$ & $\beta_0$ & $-3\alpha_0\phi/2$ & $0$ & $G_0\phi^2$ & $V_0e^{-\mu\varphi}$ & $f\big(\phi e^{\frac{\mu\varphi}{2}}\big)e^{-\mu\varphi}$   \\
\hline IV & $0$ & $\beta_0$ & $\gamma_0$ & $F_0^1+F_0\varphi$ &
$G_0^1-KF_0\phi$ &
$V_0$ & $W(\varphi-K\phi)$  \\
\hline V & $0$ & $\beta_0$ & $\gamma_0$ & $F_0$ & $G_0$ &
$V_0$ & $W(\varphi-K\phi)$  \\
\hline
\end{tabular}
\caption{Solutions.}
\end{table}

\begin{table}
\centering
\begin{tabular}{|c|c|}
\hline
Cases & $M_0$ \\
\hline
I--III & $\frac{3}{2}\alpha_0a^3\bigg\{2H\Big[4(F+G)+3\Big(\frac{2}{\mu}\frac{dF}{d\varphi}-\phi\frac{dG}{d\phi}\Big)\Big]
+\bigg(4\frac{dF}{d\varphi}-\frac{2V}{\mu\sqrt{1-\dot\varphi^2}}\bigg)\dot\varphi+\Big(4\frac{dG}{d\phi}+\phi\Big)\dot\phi\bigg\}$  \\
\hline
IV--V &
  $\gamma_0a^3\bigg\{6H\Big(K\frac{dF}{d\varphi}+\frac{dG}{d\phi}\Big)
  -\frac{KV\dot\varphi}{\sqrt{1-\dot\varphi^2}}-\dot\phi\bigg\}$  \\
\hline
\end{tabular}
\caption{Constants of motion.}
\end{table}

\subsection{Field equations and energy exchange}

From the Euler-Lagrange equations applied to the Lagrangian (\ref{plLt}) for
$a$, $\varphi$ and $\phi$, respectively, one has
\ben\no
2(F+G)(2\dot
H+3H^2)-V\sqrt{1-\dot\varphi^2}+\frac{1}{2}\dot\phi^2-W\\
+2\bigg(\frac{dF}{d\varphi}\ddot{\varphi}+2H\frac{dF}{d\varphi}\dot{\varphi}+
\frac{d^2F}{d\varphi^2}\dot{\varphi}^2\bigg)+2
\bigg(\frac{dG}{d\phi}\ddot{\phi}+2H\frac{dG}{d\phi}\dot{\phi}+
\frac{d^2G}{d\phi^2}\dot{\phi}^2\bigg)=0,\label{aet}
\\
\frac{\ddot\varphi}{1-\dot\varphi^2}+3H\dot\varphi+\frac{1}{V}\frac{dV}{d\varphi}
+\bigg[\frac{\partial
W}{\partial\varphi}-6(\dot
H+2H^2)\frac{dF}{d\varphi}\bigg]\frac{\sqrt{1-\dot\varphi^2}}{V}=0,\label{phiet}
\\
\ddot\phi+3H\dot\phi-6(\dot H+2H^2)\frac{dG}{d\phi}+\frac{\partial
W}{\partial\phi}=0.\label{chiet}
\een
{As in the previous section, by imposing} that the energy function
$E_\mathcal{L}=\frac{\partial \mathcal{L}}{\partial \dot{a}}
\dot{a}+\frac{\partial
\mathcal{L}}{\partial\dot{\varphi}}\dot{\varphi}+\frac{\partial
\mathcal{L}}{\partial\dot{\phi}}\dot{\phi}-\mathcal{L}$ associated
with the Lagrangian (\ref{plLt}) is null, it follows
\ben
6(F+G)H^2-\frac{\rho_m^0}{a^3}-\frac{V}{\sqrt{1-\dot\varphi^2}}+6H\frac{dF}{d\varphi}\dot\varphi
-\frac{1}{2}\dot\phi^2-W+6H\frac{dG}{d\phi}\dot\phi=0.\label{fet}
\een

{From the equations} (\ref{aet})-(\ref{fet}) one defines
$\rho=\rho_m+\rho_\varphi+\rho_\phi$ and $p=p_\varphi+p_\phi$, with
their forms given by
\ben
\rho_\varphi=\frac{V}{\sqrt{1-\dot\varphi^2}}-6H\frac{dF}{d\varphi}\dot\varphi,\quad\label{phidet}
\\
\rho_\phi=\frac{1}{2}\dot\phi^2+W-6H\frac{dG}{d\phi}\dot\phi,\quad\label{chidet}
\\
p_\varphi=-V\sqrt{1-\dot\varphi^2}+2\bigg(\frac{dF}{d\varphi}\ddot{\varphi}+2H\frac{dF}{d\varphi}\dot{\varphi}+
\frac{d^2F}{d\varphi^2}\dot{\varphi}^2\bigg),\quad\label{phipt}
\\
p_\phi=\frac{1}{2}\dot\phi^2-W+2\bigg(\frac{dG}{d\phi}\ddot{\phi}+2H\frac{dG}{d\phi}\dot{\phi}+
\frac{d^2G}{d\phi^2}\dot{\phi}^2\bigg),\quad\label{chipt}
\een
in {agreement with} the definitions of the energy-momentum tensors
(\ref{et2}) and (\ref{et3}).

{By using the definitions} of the energy densities (\ref{phidet}) and (\ref{chidet}) and
their respective pressures  (\ref{phipt}) and (\ref{chipt}), one has
\ben
\dot\rho_{\varphi}+3H(\rho_{\varphi}+p_{\varphi})=-\frac{\partial
W}{\partial\varphi}\dot\varphi+\frac{(dF/d\varphi)\dot\varphi}{F+G}\rho,\label{etephit}
\\
\dot\rho_{\phi}+3H(\rho_{\phi}+p_{\phi})=\frac{\partial
W}{\partial\varphi}\dot\varphi+\frac{(dG/d\phi)\dot\phi}{F+G}\rho,\label{etechit}
\een
where (\ref{phiet}), (\ref{chiet}) and (\ref{fet})
were used for the simplifications. Then, proceeding as in the
section 2, we have the covariant derivative of the total
energy-momentum tensor
\begin{equation}
\nabla_\mu
T^{\mu\nu}=\frac{\rho}{F+G}\bigg(\frac{dF}{d\varphi}\dot\varphi+\frac{dG}{d\phi}\dot\phi\bigg),\label{temtt}
\end{equation}
which has the same form of (\ref{temt}).

From these results, we will consider the interacting dark sector model
as before: one takes the field $\varphi$ to represent the dark
energy and the field $\phi$ to represent the dark matter. And
following the astronomical constrains: (i) The field
$\varphi$ composes the most part of the total energy density and has
an expressive negative pressure in the late time; (ii) The field
$\phi$ has a small positive pressure and its energy density
represents a considerable fraction of the total energy density in
the present.

\subsection{Cosmological solutions}

{Heaving in view the difficulties of integration}, we will search for numerical solutions {for the system}
(\ref{aet})-(\ref{fet}). In order to analyze the
cosmological scenarios that these models can describe, the solutions
for some cases {in Table 4} will be considered.

Let us firstly transform the derivatives with respect {to
time} in the system (\ref{aet})-(\ref{fet}) into derivatives with
respect {to red-shift}. In addition, by substituting $H^2$ from
the equation (\ref{fet}) into the equation (\ref{aet}), we obtain
the following final system of coupled differential equations to
solve
\ben
4\widetilde{H}\widetilde{H}'(1+z)(F+G)=\widetilde{\rho}+\widetilde{p},\qquad\qquad\label{ae1t}
\\\no
\frac{\widetilde{H}^2(1 + z)^2\widetilde{\varphi}''+\widetilde{H}(1
+ z)[\widetilde{H}'(1 +
z)+\widetilde{H}]\widetilde{\varphi}'}{1-\widetilde{H}^2(1
+z)^2\widetilde{\varphi}'^2}
+\frac{1}{\widetilde{V}}\frac{d\widetilde{V}}{d\widetilde{\varphi}}-3\widetilde{H}^2(1 +
z)\widetilde{\varphi}'
\\
+\bigg\{
6\widetilde{H}[\widetilde{H}'(1+z)-2\widetilde{H}]\frac{dF}{d\widetilde{\varphi}}
{+}\frac{\partial\widetilde{W}}{\partial\widetilde{\varphi}}\bigg\}
\frac{\sqrt{1-\widetilde{H}^2(1
+z)^2\widetilde{\varphi}'^2}}{\widetilde{V}}=0,\qquad\qquad\label{phie1t}
\\
\widetilde{H}^2(1 + z)^2\phi''+\widetilde{H}[\widetilde{H}'(1 +
z)-2\widetilde{H}]\bigg[(1+z)\phi'+6\frac{dG}{d\phi}\bigg]+\frac{\partial\widetilde{W}}{\partial\phi}=0,\label{chie1t}
\een
with the line representing derivative with respect to $z$, where
$\widetilde{H}=H/\sqrt{\rho_0}$,
$\widetilde{\varphi}=\sqrt{\rho_0}\varphi$,
$\widetilde{V}=V/\rho_0$, $\widetilde{W}=W/\rho_0$,
$\widetilde{\rho}=\rho/\rho_0=\rho_m/\rho_0+\rho_\varphi/\rho_0+\rho_\phi/\rho_0=\widetilde{\rho}_m+
\widetilde{\rho}_{\widetilde{\varphi}}+\widetilde{\rho}_\phi$ and
$\widetilde{p}=p/\rho_0=p_\varphi/\rho_0+p_\phi/\rho_0=\widetilde{p}_{\widetilde{\varphi}}+\widetilde{p}_\phi$,
which are dimensionless quantities. The energy densities and
pressures are now given by
\ben
\widetilde{\rho_m}=\widetilde{\rho}_m^0(1 + z)^3,\label{edm1t}
\\
\widetilde{\rho}_{\widetilde{\varphi}}=\frac{\widetilde{V}}{\sqrt{1-\widetilde{H}^2(1
+ z)^2\widetilde{\varphi}'^2}}+6\widetilde{H}^2(1 +
z)\frac{dF}{d\widetilde{\varphi}}\widetilde{\varphi}',\label{edphi1t}
\\
\widetilde{\rho}_{\phi}=\frac{\widetilde{H}^2(1 +
z)^2\phi'^2}{2}+\widetilde{W}+6\widetilde{H}^2(1 +
z)\frac{dG}{d\phi}\phi',\label{edchi1t}
\\
\widetilde{p}_{\widetilde{\varphi}}=2\widetilde{H}(1+z)\bigg\{\widetilde{H}(1+z)
\bigg(\frac{d^2F}{d\widetilde{\varphi}^2}\widetilde{\varphi}'^2+
\frac{dF}{d\widetilde{\varphi}}\widetilde{\varphi}''\bigg)+[\widetilde{H}'(1+z)-\widetilde{H}]\frac{dF}{d\widetilde{\varphi}}\widetilde{\varphi}'\bigg\}
\nonumber\\
-\widetilde{V}\sqrt{1-\widetilde{H}^2(1
+z)^2\widetilde{\varphi}'^2},\label{pphi1t}
\\
\widetilde{p}_{\phi}=2\widetilde{H}(1+z)\bigg\{\widetilde{H}(1+z)\bigg(\frac{d^2G}{d\phi^2}\phi'^2+
\frac{dG}{d\phi}\phi''\bigg)+[\widetilde{H}'(1+z)-\widetilde{H}]\frac{dG}{d\phi}\phi'\bigg\}
\nonumber\\
+\frac{\widetilde{H}^2(1 +
z)^2\phi'^2}{2}-\widetilde{W}.\label{pchi1t}
\een

The requirements (i) and (ii) {for the fields} $\phi$ and $\varphi$ will be
satisfied by using the initial conditions for the system
(\ref{ae1t})-(\ref{chie1t}) {determined} from the astronomical data, as it is
done in the {canonical - canonical model}. From the requirement I we
have that $\widetilde{\varphi}'(0)^2\ll1$, that is, the field
$\varphi$ is varying very slowly in the late time (the same consideration of Section 2). This condition
and  (\ref{edphi1t}) {imply} in the relation
$\widetilde{V}(0)\approx\Omega_\varphi^0$. Remembering that the
gravitational coupling must present the value 1/2 in the present,
$F(0)+G(0)=1/2$, the equation (\ref{fet}) furnishes the initial
condition $\widetilde{H}(0)=\sqrt{\Omega_0/6[F(0)+G(0)]}=1/\sqrt{3}$
to the Hubble parameter, just as it was in the first case. All these
relations will be employed to perform comparisons to the
observational data.

From now on, we will analyze the {cases I and V} from
{Table 4}, which represent interacting models non-minimally and
minimally coupled to the gravity, respectively.

For the case I,  we have chosen
\begin{equation}\nonumber f\left(\phi
e^{\frac{\mu\varphi}{2}}\right)={e^{-\frac{\mu\varphi}{2}}\over\phi},\quad\hbox{which
implies}\quad W=W_0{e^{-\frac{3\mu\varphi}{2}}\over\phi}.
\end{equation} For theses functions, one may determine that the initial
conditions are given by:
\ben\nonumber
\widetilde{\varphi}(0)={\ln[\widetilde{V_0}/0.72]\over\widetilde{\mu}},\qquad
\phi(0)=\sqrt{{1/2-F_0e^{-\widetilde{\mu}\widetilde{\varphi}(0)}\over
G_0}},\qquad
\\ \nonumber\phi'(0)=\sqrt{6\left[0.23+24G_0^2\phi(0)^2-\widetilde{W_0}\phi(0)^{-1}e^{-\frac{3\widetilde{\mu}
\widetilde{\varphi}(0)}{2}}\right]}-12G_0\phi(0),
\een
 where $\widetilde{\mu}=\mu/\sqrt{\rho_0}$ and
 \begin{equation}\nonumber F_0\leq \frac{e^{\widetilde{\mu}\widetilde{\varphi}(0)}}{2},\qquad
\widetilde{W}_0\leq\big[0.23+24G_0^2\phi(0)^2\big]\phi(0)e^{\frac{3\widetilde{\mu}\widetilde{\varphi}}{2}}.
 \end{equation}

For the derivative of the field ${\varphi}$ at $z=0$ we have chosen $\widetilde{\varphi}'(0)=10^{-6}$ and the following values  have been adopted for the fixed constants: $F_0 = -8.5\times10^{-3}; G_0 = 6.9\times10^{-3}; \widetilde{V_0} =2.2\times10^{-3}$; $\widetilde{W_0} = 1.9\times10^{-3}$.  Two values for the coefficient in the exponential term were adopted, namely, $\widetilde{\mu_1}=10^{-3}$ and $\widetilde{\mu_2}=10^{-2}$.

For the case V we have considered that the pressure of the dark matter vanishes  at $z=0$ and that the {interaction potential} of the scalar fields is given by
$$W(\varphi-\kappa\phi)=W_0e^{-\xi(\varphi-\kappa\phi)}.$$
{From the subtraction and sum of the equations (\ref{edchi1t}) and (\ref{pchi1t}) one obtains the initial conditions for $\phi(0)$ and $\phi'(0)$, respectively,}
$$\phi(0)={\widetilde{\varphi}(0)+\ln{\left(0.115/\widetilde{W_0}\right)}/\widetilde{\xi}\over\widetilde{\kappa}},\qquad \phi'(0)=\sqrt{0.69},$$
where $\widetilde{\xi}=\xi/\sqrt{\rho_0}$ and $\widetilde{\kappa}=\sqrt{\rho_0}\kappa$. The initial conditions for $\widetilde{\varphi}(0)$ and $\widetilde{\varphi}'(0)$  are free and were chosen as $\widetilde{\varphi}(0)=1.0$ and $\widetilde{\varphi}'(0)=10^{-2}$. For the fixed constants {the following values were adopted}: $F_0+G_0=1/2$; $\widetilde{V_0} = 0.72$; $\widetilde{W_0}=10^{-2}$; $\widetilde{\kappa}=0.5405$. As in the previous case, two values for the coefficient in the exponential term were adopted, namely,  $\widetilde{\xi_1}=4.9$ and $\widetilde{\xi_2}=4.45$.

\begin{figure}[ht]
\begin{center}
\vskip0.5cm
\includegraphics[height=4.7cm,width=6.7cm]{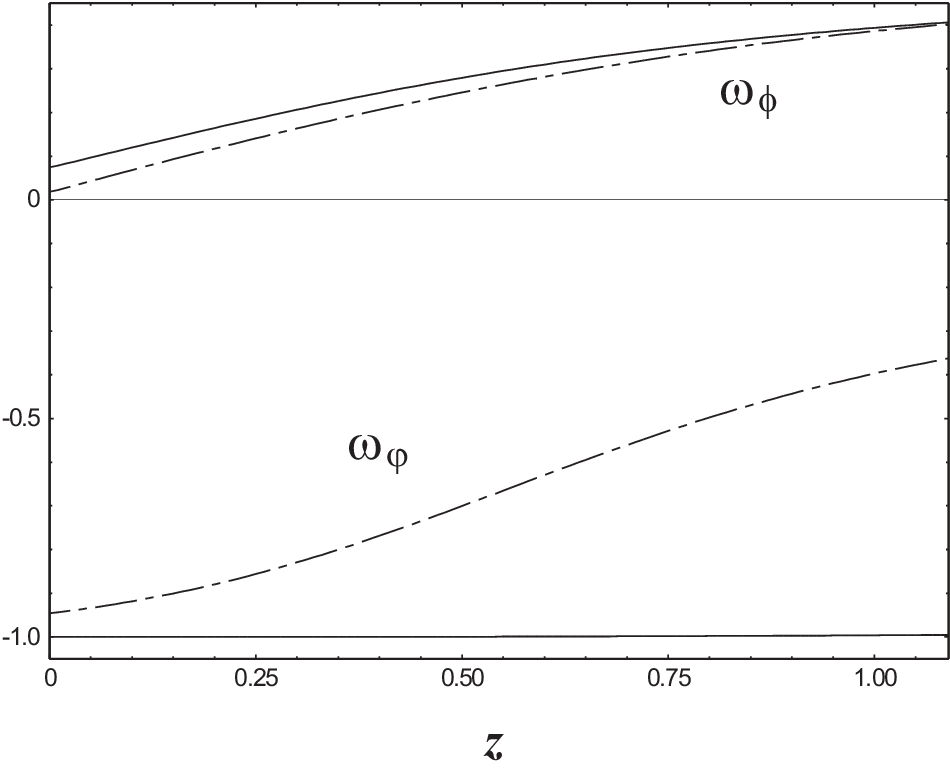}\hskip0.5cm
\includegraphics[height=4.7cm,width=6.5cm]{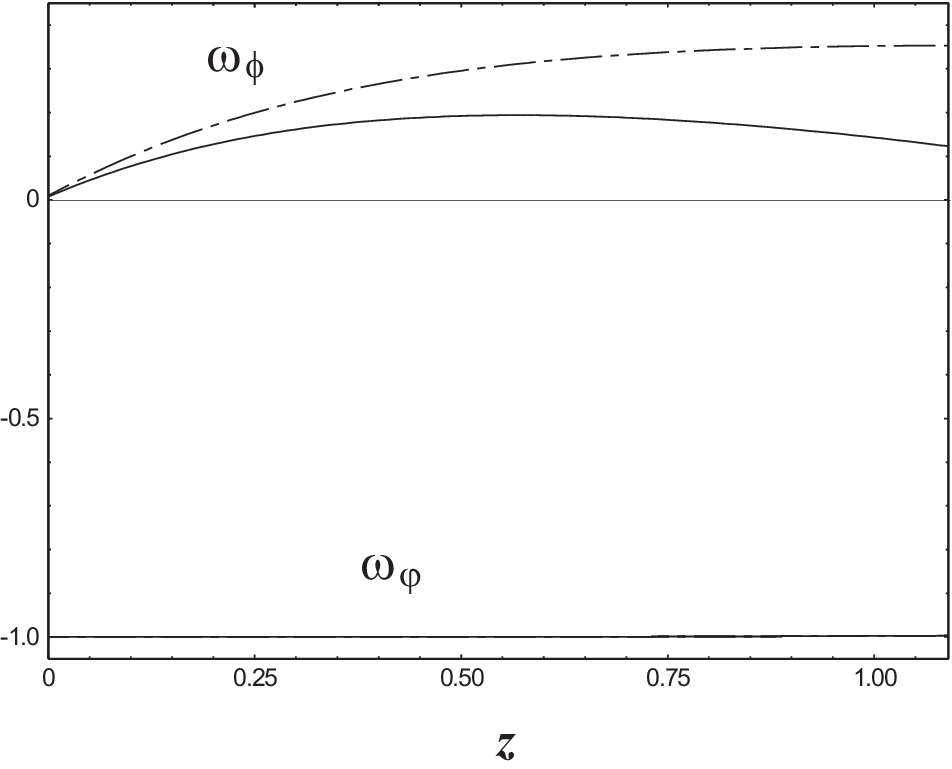}
\caption{Left frame: the ratio of the pressure and energy density of
the scalar fields for the case I, represented by the straight line
for {$\widetilde{\mu_1}=10^{-3}$} and by the dashed line for
{$\widetilde{\mu_2}=10^{-2}$}. Right frame: the ratio of the pressure
and energy density of the scalar fields for the case V, represented
by the straight line for $\widetilde{\xi_1}=4.900$ and by the dashed
line for $\widetilde{\xi_2}=4.450$.}
\end{center}
\end{figure}

In Figure 4 are plotted the ratio of the pressure and energy density of the scalar fields, $\omega_\varphi={p_\varphi}/{\rho_\varphi}$ and ${\omega}_\phi={p_\phi}/{\rho_\phi}$, where the left and right frames represent the cases I and V, respectively. From this figure one can infer that when  the parameter $\widetilde{\mu}$ is varied from $10^{-3}$ to $10^{-2}$ (case I) the ratio $\omega_\varphi$ changes its red-shift evolution drastically. This behavior can be understood because -- {according to  (\ref{etephit})} -- this ratio
 is related to the direct exchange of energy {between the field $\varphi$ and gravitational field}. Then the behavior of the dark energy changes from a cosmological constant-type $\omega_\varphi\approx-1$ for $\widetilde{\mu_1}=10^{-3}$ to a variable $\omega_\varphi$  for $\widetilde{\mu_2}=10^{-2}$ as a consequence {of the modification} in the direct energy exchange with the gravitational field. However, the ratio $\omega_\phi$ has a smooth variation when the values of the coefficient in the exponential term are changed. For the case V, when the parameter $\widetilde{\xi}$ is varied, one notes by observing the behavior of the ratio $\omega_\phi$ that the dark matter suffers a significant influence, while the dark energy always has an approximated cosmological constant-type behavior, since a very small variation of the ratio $\omega_\varphi$ {occurs}. Observe that in the case V there is no direct energy exchange with the gravitational field  due to the conditions $\{F, G\}=$ constant.
\begin{figure}[ht]
\begin{center}
\vskip0.5cm
\includegraphics[height=4.7cm,width=6.7cm]{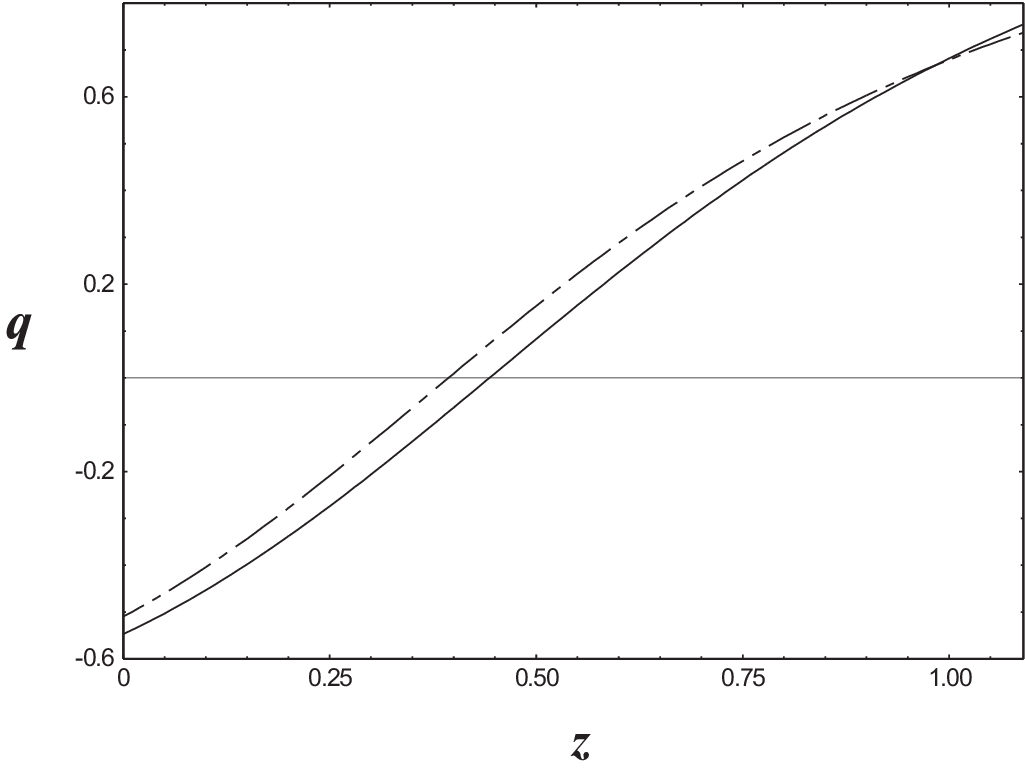}\hskip0.5cm
\includegraphics[height=4.7cm,width=6.5cm]{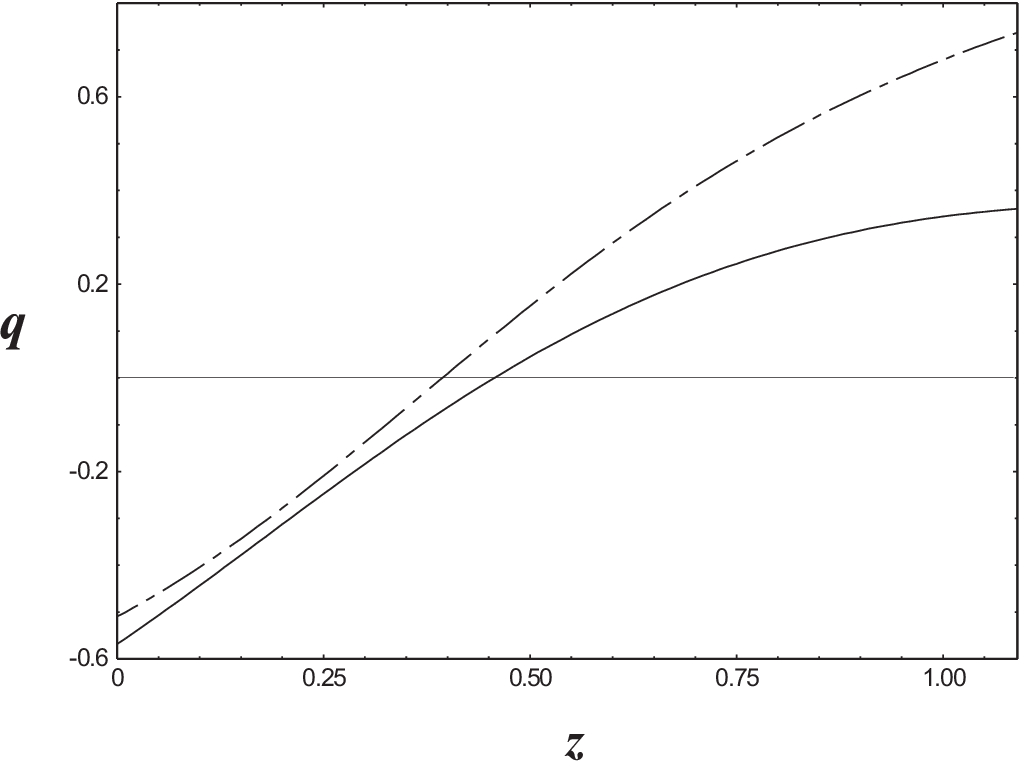}
\caption{Left frame: deceleration parameter for the case I,
represented by the straight line for {$\widetilde{\mu_1}=10^{-3}$}
and by the dashed line for {$\widetilde{\mu_2}=10^{-2}$}. Right frame:
deceleration parameter for the case V, represented by the straight
line for $\widetilde{\xi_1}=4.900$ and by the dashed line for
$\widetilde{\xi_2}=4.450$.}
\end{center}
\end{figure}

\begin{figure}[ht]
\begin{center}
\vskip0.5cm
\includegraphics[height=4.7cm,width=6.7cm]{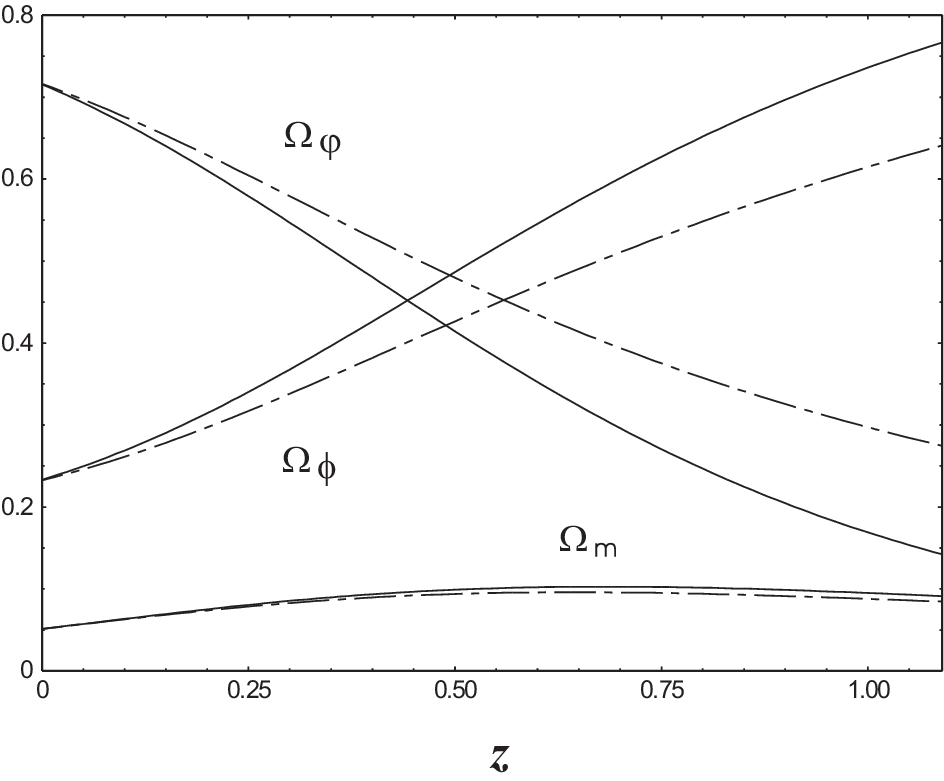}\hskip0.5cm
\includegraphics[height=4.7cm,width=6.5cm]{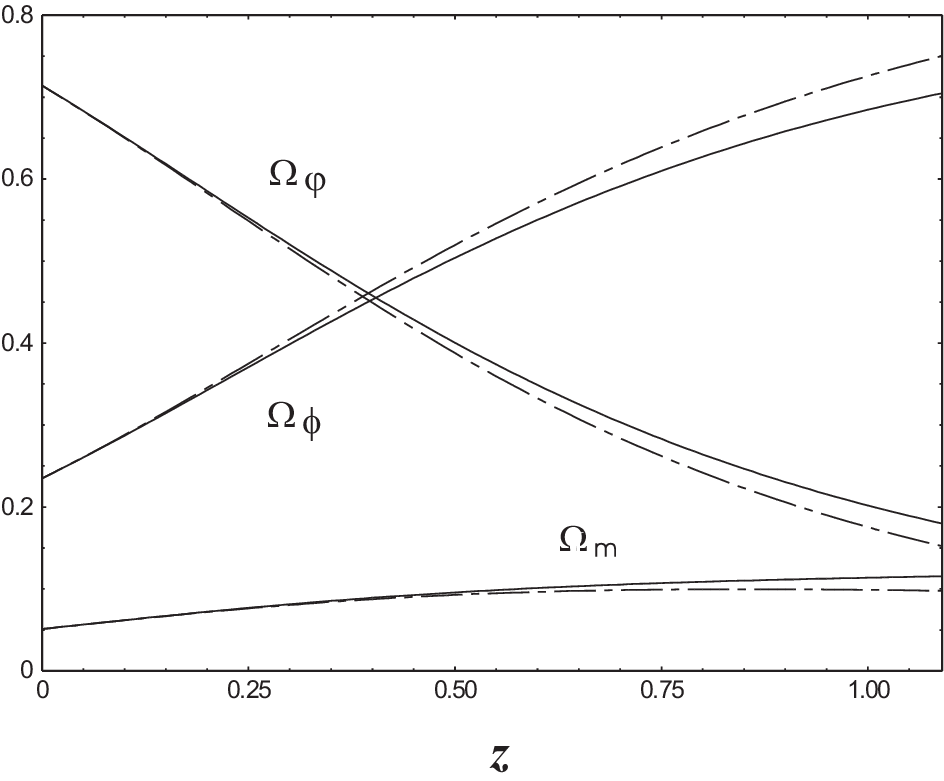}
\caption{Left frame: parameters of density for the case I,
represented by the straight line for {$\widetilde{\mu_1}=10^{-3}$}
and by the dashed line for {$\widetilde{\mu_2}=10^{-2}$}. Right frame:
parameters of density for the case V, represented by the straight
line for $\widetilde{\xi_1}=4.900$ and by the dashed line for
$\widetilde{\xi_2}=4.450$.}
\end{center}
\end{figure}

\begin{figure}[ht]
\begin{center}
\vskip0.5cm
\includegraphics[height=4.7cm,width=6.7cm]{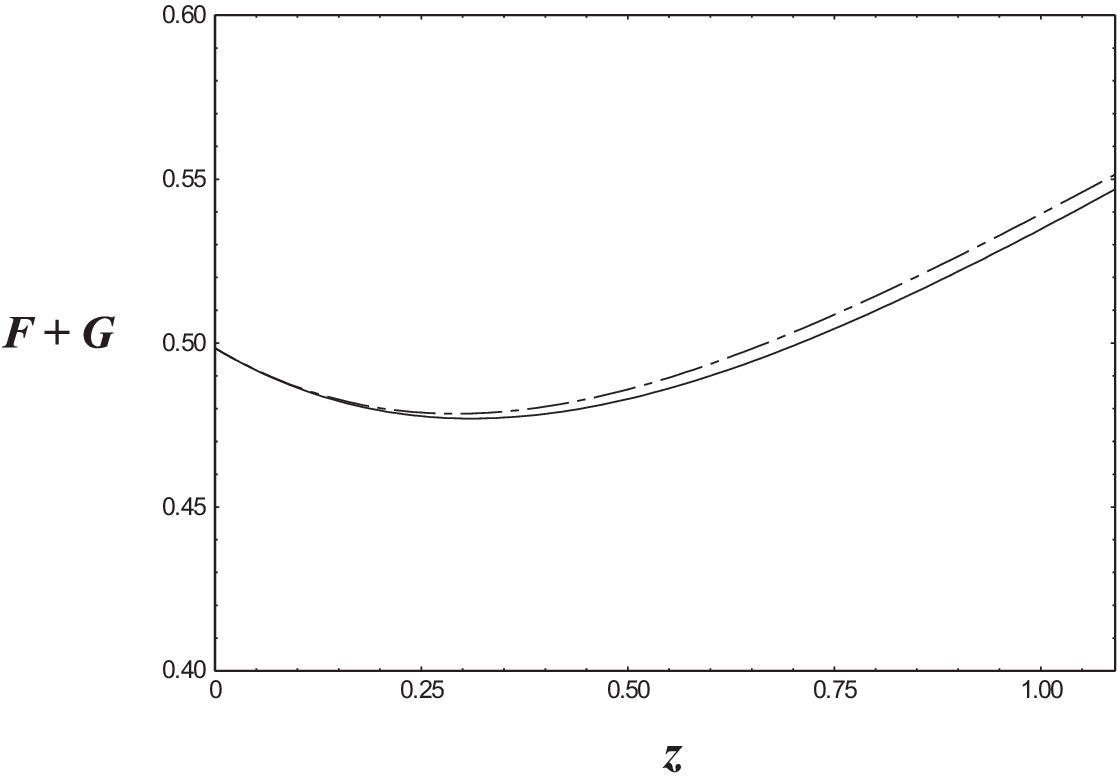}
\caption{Effective gravitational coupling for the case I,
represented by the straight line for {$\widetilde{\mu_1}=10^{-3}$}
and by the dashed line for {$\widetilde{\mu_2}=10^{-2}$}.}
\end{center}
\end{figure}

 The deceleration parameter $q = 1/2+{3p}/{2\rho}$ is represented in  Figure 5, for the cases I (left frame) and V (right frame). The left frame of this figure shows us that the deceleration parameter exhibits a small modification when one varies the value of the {coefficient $\widetilde{\mu}$ in the exponential term}. However, one may infer that when $\widetilde{\mu}_2=10^{-2}$ and   $\omega_\varphi$  goes asymptotically to $-1$ in the late time, the red-shift transition from a decelerated to a accelerated regime is smaller than that for $\widetilde{\mu}_1=10^{-3}$. {Indeed, in this last situation $\omega_\varphi\approx-1$ in the whole evolution of the density parameter, which implies an earlier transition of regime.} For the case V, one may  observe that for $\widetilde{\xi_2}=4.45$ the red-shift transition is smaller than that for $\widetilde{\xi_1}=4.9$. This can be understood by looking {at Figure 4} again, where one can infer that $\omega_\phi$ is larger for $\widetilde{\xi_2}=4.45$  than for $\widetilde{\xi_1}=4.9$, meaning that in this situation the dark matter has a larger relative pressure, which contributes to  retard  the {transition of regime}. The values of the {red-shift transition} for the case I are: $z_T = 0.45$ ($\widetilde{\mu}_1=10^{-3}$) and $z_T = 0.40$ ($\widetilde{\mu}_2=10^{-2}$), whereas those for the case V are: $z_T = 0.46$ ($\widetilde{\xi}_1=4.9$) and $z_T = 0.40$ ($\widetilde{\xi}_2=4.45$). Furthermore, the values of the deceleration parameter $q_0$ at $z=0$  for the case I are: $q_0 = -0.55$ ($\widetilde{\mu}_1=10^{-3}$) and $q_0 = -0.51$ ($\widetilde{\mu}_2=10^{-2}$),  while those for the case V are: $q_0 = -0.57$ ($\widetilde{\xi}_1=4.9$) and $q_0 = -0.51$ ($\widetilde{\xi}_2=4.45$). {In order to perform comparisons to the observational data, the recent observed values are $z_T = 0.74 \pm 0.18$ (from \cite{26}) for the red-shift  transition and $q_0 = -0.46 \pm 0.13$ (from \cite{27}) for the deceleration parameter at $z=0$}. Hence, one may conclude that there exists a good agreement of the results with the observational data.

The density parameters of the {common matter}, dark energy and dark matter fields are represented in  Figure 6 for the cases I (left frame) and V (right frame). From the left frame one observes that for $\widetilde{\mu_2}$ {the density parameters of the dark energy and dark matter become equals earlier} {than for} $\widetilde{\mu_1}$, but its red-shift transition occurs at a smaller red-shift than that for $\widetilde{\mu_1}$. This shows that the change of the behavior of $\omega_\varphi$ caused by the change of the energy exchange with the gravitational field really is the responsible for a smaller red-shift transition in the case I. On the other hand, by observing the right frame, one notes that {for $\widetilde{\xi_1}$ the density parameters of the dark energy and dark matter become equals earlier than for} $\widetilde{\xi_2}$. {This reinforces the delay of the red-shift transition caused by a larger relative pressure of the dark matter for $\widetilde{\xi_2}$}.

The effective coupling $F+G$ for the case I is plotted in  Figure 7.
One observes from this figure that the effective gravitational coupling has a small variation in comparison to its present value $F(0)+G(0)$ = 1/2. {There is} a variation of less than 10\% around the value 1/2, and consequently the effective gravitational "constant" has its value changed about 10\% in the considered interval. {This result is similar to that of the canonical - canonical case.}

\section{Conclusions}

By applying the Noether symmetry approach we have restricted the possible functions of the undefined couplings and potentials of the general models to families of functions. Using this tool we could analyze the cosmological solutions of some particular interacting dark sector models, which correspond to potentials and couplings
satisfying the symmetry condition for the general actions. Some of the resulting models from the symmetry condition generalize certain interacting dark sectors models that have appeared in the literature.
The energy exchange which occurs among the fields (scalar field -
scalar field and gravitational field - scalar fields) has strong
influence on the behaviors described by models with scalar fields.
An important verification is that non-minimal couplings can have a very significant {influence on the} evolution of the energy densities and pressures of the components of the Universe.
The results for both general models ({canonical - canonical} and non-canonical - canonical) showed distinct ways to the energy density and pressure evolutions in similar regimes of expansion of the Universe. Further, both can reproduce a decelerated-accelerated regime, describing the recent transition from a decelerated to an accelerated expansion in {agreement with} the observational data. {Canonical - non-canonical models can reproduce behaviors very similar to that of the cosmological constant model to the late Universe (with respect to the ratio of the pressure and energy density of the dark energy field), but with the additional advantage of presenting more possible ways for the energy density evolution of the matter fields.}

\end{document}